\documentclass[twocolumn,pra,aps,superscriptaddress,showpacs]{revtex4}
\usepackage{graphicx}
\begin{document}
\title {Cold-atoms on a two-dimensional square optical lattice with
an alternating potential}
\author{Huaiming Guo}
\author{Yuchuan Wen}
\affiliation{Department of Physics, Capital Normal University,
Beijing 100048, China}
\author{Shiping Feng}
\affiliation{Department of Physics, Beijing Normal University,
Beijing 100875, China}

\begin{abstract}
The cold-atom on a two-dimensional square optical lattice is studied
within the hard-core boson Hubbard model with an alternating
potential. In terms of the quantum Monte Carlo method, it is shown
explicitly that a supersolid phase emerges due to the presence of
the alternating potential. For the weak alternating potential, the
supersolid state appears for the whole range of hard-core boson
densities except the half-filling case, where the system is a Mott
insulator. However, for the strong alternating potential, besides
the supersolid and Mott insulating states, a charge density wave
phase appears.
\end{abstract}
 \maketitle

The optical lattice, which is produced by using laser beams, allows
one to tune relevant parameters (such as the confining potential,
particle density, interaction, lattice geometry, lattice constant,
etc.), and therefore to make it an ideal test-ground for concepts
and theories on condensed matter physics, especially strongly
correlated many-body quantum lattice models
\cite{banderson,corzel,omorsch,djaksch,lguidoni}. Theoretically, it
has been argued that the physical properties of bosons trapped in an
optical lattice can be described within the boson Hubbard model.
This boson Hubbard model and its extended version have been studied
extensively
\cite{ggbatouni,swessel,dheidarian,rgmelko,mboninsegni,jygan,psengupta},
and the results show a number of interesting features, such as the
existence of superfluid state, charge density wave (CDW) state, and
Mott insulating state.

Over fifty years ago, Penrose and Onsager \cite{npensrose} suggested
a novel state with simultaneous solid and superfluid phases, which
intrigued in condensed matter physics field. We call such state the
supersolid state. Recently, a possible supersolid state in the solid
$^4$He has renewed interests in this new matter state \cite{ekim}.
However the cold-atom in an optical lattice is a more promising
realization of supersolid phase. Theoretical studies
\cite{ggbatouni,rgmelko,mboninsegni,jygan,swessel,dheidarian,psengupta}
have shown that next-nearest-neighbor interaction, frustration of
lattice, and softening of the on-site interaction are favorable for
the realization of the supersolid state. However, based on the
two-dimensional (2D) hard-core boson Hubbard model on a square
lattice with nearest-neighbor repulsion interaction, it has been
found that there is no supersolid phase due to the solid-superfluid
phase separation \cite{ggbatouni}. Motivated by the ionic Hubbard
model \cite{tegami}, whether or not the phase diagram of the 2D
hard-core boson Hubbard model can be influenced by an alternating
potential is an exciting issue.

In this paper, we study this important problem. Within the hard-core
boson Hubbard model on a square lattice with an alternating
potential, we employ the quantum Monte Carlo (QMC) method, and show
explicitly that the supersolid phase emerges due to the presence of
the alternating potential. In particular, our results also indicate
that even very weak alternating potential can prevent the occurrence
of the solid-superfluid phase separation, and then induce the
supersolid state.

The 2D hard-core boson Hubbard model on a square lattice with an
alternating potential can be expressed as
\begin{eqnarray}
H=-t\sum_{<ij>}(a_{i}^{\dagger}a_{j}+a_{j}^{\dagger}a_{i})-
\mu\sum_{i}\hat{n}_{i}+\Delta\sum_{i}(-1)^{i_{x}+i_{y}}\hat{n}_{i},
\end{eqnarray}
where $<ij>$ indicates the sum over nearest neighbor sites, the
operator $a_{i}^{\dagger}$ ($a_{i}$) creates (destroys) a hard-core
boson on site $i$, $\hat{n}_{i}=a_{i}^{\dagger}a_{i}$ is the
hard-core boson number operator, $\mu$ is the chemical potential and
therefore controls the filling number in the trap, hopping $t$
describes the coherent hopping between nearest-neighbor sites, and
$(-1)^{(i_{x}+i_{y})}\Delta$ is the alternating potential. In the
hard-core limit, the Hubbard constant $U\rightarrow\infty$. In this
case, each lattice site can be occupied by 0 or 1 hard-core boson.
Since hard-core bosons are restricted in this Hilbert subspace,
operators $a_{i}^{\dagger}$ and $a_{i}$ obey commutation relations
$[a_{i},a_{j}^{\dagger}]=0$ at sites $i\ne j$, while they satisfy
anti-commutation relations $\{a_{i},a_{i}^{\dagger}\}=1$ on sole
site $i$.

In the following discussions, our QMC simulation is based on the
stochastic series expansion method with directed loop updates
\cite{sandvik}. In Hamiltonian (1), the different phases are
characterized by static staggered structure factor $S({\bf Q})$ with
wave vector ${\bf Q}=[\pi,\pi]$ and superfluid density $\rho_{s}$,
where the structure factor measures the diagonal long-range order of
the system, and the superfluid density measures the off-diagonal
long-range order. We emphasize that in the square lattice magnetic
system, the wave vector ${\bf Q}$ is an antiferromagnetic wave
vector. However, in the present square lattice hard-core boson
system, this wave vector ${\bf Q}$ is similar to that in the
magnetic system, and therefore the structure factor $S({\bf Q})$
measures the diagonal long-range order with the checkerboard
pattern. In the present case, the structure factor, the superfluid
density, and the hard-core boson density are expressed as
\cite{psengupta,elpollock}:
\begin{eqnarray}
\rho &=&{1\over N}\sum_{i}\langle\hat{n}_{i}\rangle={1\over N}
\sum_{i}\langle a_{i}^{\dagger}a_{i} \rangle,\\
 S({\bf Q})&=&{1\over
N}\sum_{jj'}e^{i{\bf Q}\cdot({\bf R}_{j}-{\bf
R}_{j'})}\langle \hat{n}_{j}\hat{n}_{j'}\rangle, \\
\rho_{s}&=&{\langle W^2\rangle\over 4\beta t},
\end{eqnarray}
where $W$ is the winding number of the bosonic world lines. Due to
the presence of the alternating potential $(-1)^{(i_{x}+i_{y})}
\Delta$ in Hamiltonian (1), there are two sublattices A and B in the
system. This leads to the formation of two energy bands, where the
gap is opened by the alternating potential. At half-filling
(corresponding to the hard-core boson density $\rho=1$), both energy
bands are completely occupied by hard-core bosons, so all hard-core
bosons are localized, with $S({\bf Q}) =0$ and $\rho_{s}=0$, and the
corresponding state of the system is a Mott insulating state. For
the case of the hard-core boson density $\rho=1/2$, the sublattice
with minus potential is completely occupied by the hard-core bosons,
while the other sublattice is empty, and the corresponding state of
the system is CDW state, with $S({\bf Q})\neq 0$ and $\rho_{s}=0$.
The conditions for the supersolid phase are that both $S({\bf Q})
\neq 0$ and $\rho_{s}\neq 0$ simultaneously. The supersolid state is
a novel state with simultaneous diagonal (CDW solid) and
off-diagonal (superfluid) long-range orders \cite{psengupta}. In the
following QMC calculations, we discuss the structure factor $S({\bf
Q})$ and the superfluid density $\rho_{s}$ as a function of the
hard-core boson density $\rho$, the hard-core boson density $\rho$
as a function of the chemical potential $\mu$, and the ground state
phase diagram of the hard-core boson Hubbard model (1) with an
alternating potential in the $\mu-\Delta$ plane, where all the
numerical results have been performed on a $N=L\times L$ lattices
with $L=16$, and the inverse temperature was chosen as $\beta=2L$,
which is low enough for the numerical simulation of the ground state
properties of the cold-atom in an optical lattice. The finite size
extrapolation of the following measurements converges to a finite
value, indicating that the obtained phase is stable in the
thermodynamic limit.

\begin{figure} \centering
\includegraphics[width=9cm]{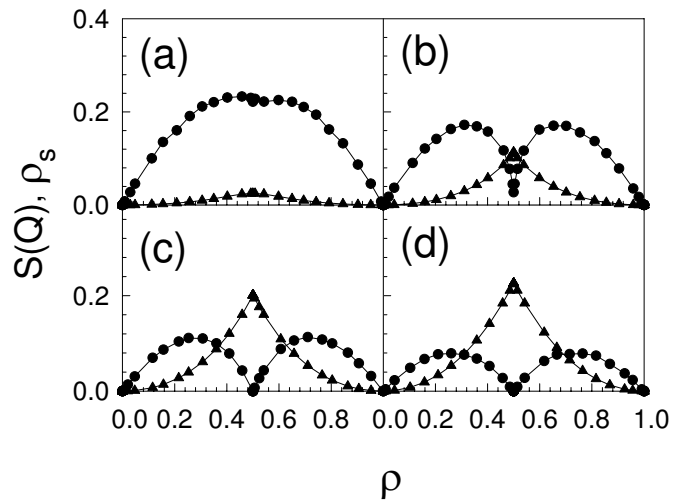}
\vspace{-0.2cm}\caption{\label{Fig1} The superfluid density
$\rho_{s}$ (circle) and structure factor $S({\bf Q})$ (triangle) as
a function of the hard-core boson density $\rho$ for (a)
$\Delta/t=1$, (b) $\Delta/t=2$, (c) $\Delta/t=4$, and (d)
$\Delta/t=6$} \label{fig1}
\end{figure}

\begin{figure} \centering
\includegraphics[width=9cm]{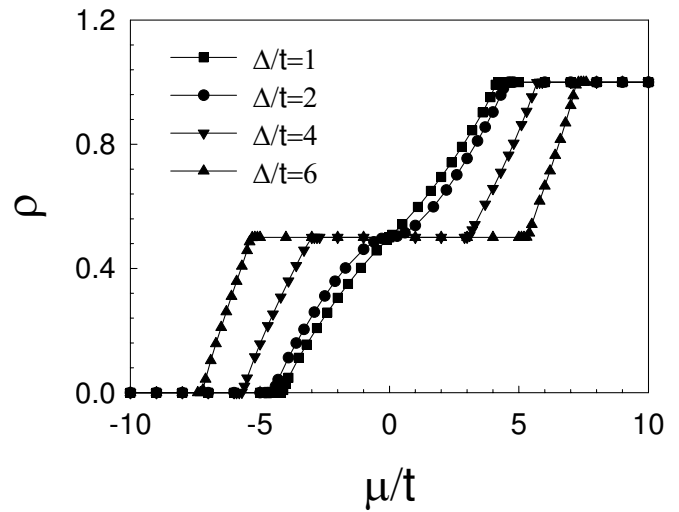}
\vspace{-0.2cm}\caption{\label{Fig2} The hard-core boson density
$\rho$ as a function of the chemical potential $\mu$ for
$\Delta/t=1$ (square), $\Delta/t=2$ (circle), $\Delta/t=4$ (triangle
down), and $\Delta/t=6$ (triangle up).} \label{fig2}
\end{figure}
Firstly, we plot the superfluid density $\rho_s$ (circle) and the
structure factor $S({\bf Q})$ (triangle) as a function of the
hard-core boson density $\rho$ for parameter (a) $\Delta/t=1$, (b)
$\Delta/t=2$, (c) $\Delta/t=4$, and (d) $\Delta/t=6$ in Fig. 1,
where the superfluid density $\rho_s$ and the structure factor
$S({\bf Q})$ are changed dramatically with the alternating potential
$\Delta$. In particular, for any alternating potential, the
superfluid density and the structure factor are symmetrical around
$\rho=1/2$, reflecting the particle-hole symmetry. Moreover, for any
alternating potential, the Mott insulating state appears only at
half-filling ($\rho=1$). Surprisingly, for the weak alternating
potential, both $S({\bf Q})\neq 0$ and $\rho_{s}\neq 0$ at
$0<\rho<1$. In this case, CDW state is absent, and the supersolid
state appears for all hard-core boson densities except $\rho=0$ and
$\rho=1$. This is much different from the case of the 2D hard-core
boson Hubbard model on a square lattice with nearest-neighbor
repulsion interaction, where the supersolid phase is
thermodynamically unstable, and therefore no supersolid phase can be
observed due to the CDW solid-superfluid phase separation
\cite{psengupta,ggbatouni}. However, for the strong alternating
potential, besides the Mott insulating and supersolid states, CDW
state appears at $\rho=1/2$. Moreover, the superfluid density
$\rho_{s}\rightarrow 0$ at $\rho=1/2$ while $S({\bf Q})\sim 1/4$,
which is an exact value of the static structure factor of the
checkerboard solid.

\begin{figure} \centering
\includegraphics[width=9cm]{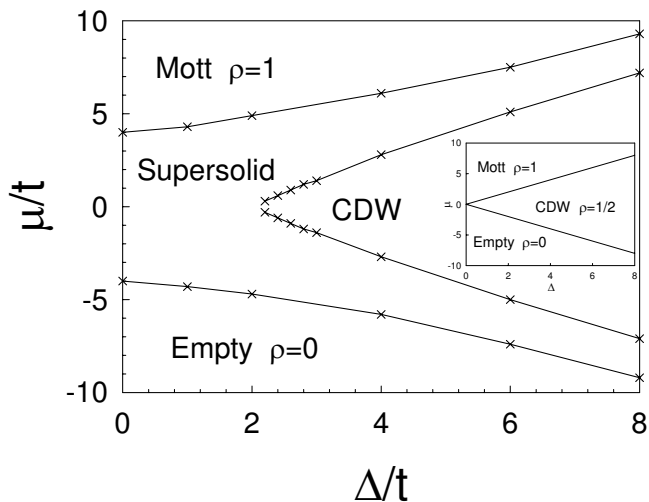}
\vspace{-0.2cm}\caption{\label{Fig3} The ground state phase diagram
in the $\mu-\Delta$ plane. Inset: the corresponding result for
$t=0$.}\label{fig3}
\end{figure}

For a further confirmation of the above obtained conclusion, we have
made a series of calculations for the hard-core boson density $\rho$
as a function of the chemical potential $\mu$, and the results for
$\Delta/t=1$ (square), $\Delta/t=2$ (circle), $\Delta/t=4$ (triangle
down), and $\Delta/t=6$ (triangle up) are plotted in Fig. 2.
Obviously, for any alternating potential, $\rho$ varies continuously
with $\mu$, reflecting a phase transition of second order. When
$\mu$ is less than $-\Delta$, hard-core bosons begin to reside in
the optical lattice due to the presence of hopping $t$. This
critical value $\mu_{c}$ obtained from QMC simulations is
quantitatively consistent with $\mu_{c}=-\sqrt{\Delta^2+16t^2}$
obtained from analytical calculations. Moreover, a plateau at
$\rho=1/2$ appears for the strong alternating potential. In
comparison with the results in Fig. 1, we therefore find that the
state in this plateau corresponds to a CDW state. This plateau
decreases with the decrease of $\Delta$, and vanishes in the weak
alternating potential. This result could be anticipated, since the
strong alternating potential favors CDW state. For the large values
of $\mu$, the system is at half-filling, so this leads to the
formation of an uniform Mott insulating state. Furthermore, the
range of the supersolid state increases with the decrease of
$\Delta$, and then in the weak alternating potential, the ground
state for all hard-core boson densities is the supersolid state
except the Mott insulating state at $\rho=1$. This is a remarkable
result, since it indicates that even very weak alternating potential
can prevent the occurrence of the solid-superfluid phase separation,
and then induce the supersolid state. To show these points clearly,
we plot the ground state phase diagram in the $\mu-\Delta$ plane in
Fig. 3. In the case of hopping $t=0$, the hard-core boson Hubbard
model (1) with an alternating potential is reduced to
$H=-\mu\sum_{i}n_{i} +\Delta\sum_{i}(-1)^{i_{x}+ i_{y}}n_{i}$. For
comparison, the phase diagram in this case ($t=0$) is also plotted
in Fig. 3 (inset). It appears that the supersolid state, which
appeares for $t\ne 0$ is not observed for $t=0$, and the ground
state is CDW state at $\rho=1/2$ or Mott insulating state at
$\rho=1$. On the other hand, in the case of the alternating
potential $\Delta=0$, the model Hamiltonian (1) is reduced to the
simple hard-core boson Hubbard model
$H=-t\sum_{<ij>}(a_{i}^{\dagger}a_{j}+ a_{j}^{\dagger}a_{i})-
\mu\sum_{i}n_{i}$, where the superfluid phase [$S({\bf Q})=0$ and
$\rho_{s}\ne 0$] appears at $0<\rho<1$ \cite{psengupta}, and no
supersolid state is observed. In this case, the supersolid phase in
Fig. 3 does not include the line of $\Delta=0$. These results
therefore confirm that for finite hopping $t$, the supersolid state
emerges due to the presence of the alternating potential in the
hard-core boson Hubbard model, in particular, for the weak
alternating potential $0<\Delta/t<2$, the ground state for all
hard-core boson densities is the supersolid state except the Mott
insulating state at $\rho=1$ and the empty state at $\rho=0$.

\begin{figure} \centering
\includegraphics[width=9cm]{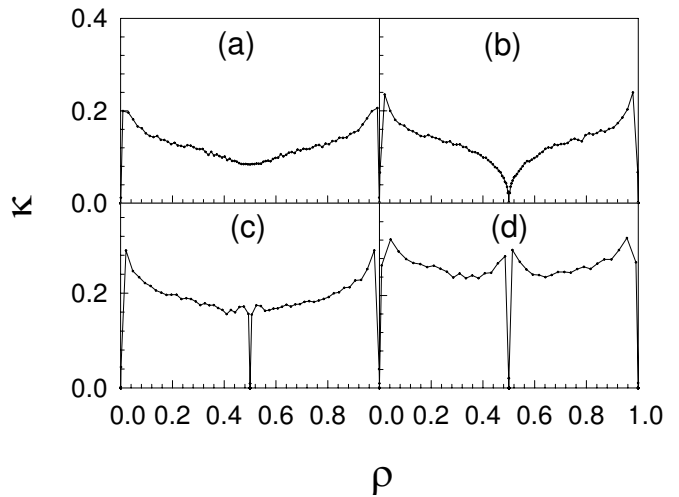}
\vspace{-0.2cm}\caption{\label{Fig3} The compressibility $\kappa$ as
a function of the hard-core boson density $\rho$ for (a)
$\Delta/t=1$, (b) $\Delta/t=2$, (c) $\Delta/t=4$, and (d)
$\Delta/t=6$, where the unit of $\kappa$ is 1/t.} \label{fig4}
\end{figure}

One of the characteristic features for the thermodynamically stable
state is that it has a positive compressibility. To better perceive
the stable states shown in Fig. 3, we turn now to the discussion of
the compressibility of the system, which is defined by $\kappa=\beta
N\langle\rho^{2}-\langle\rho\rangle^{2} \rangle$ \cite{min}. We have
performed a QMC simulation for the compressibility $\kappa$, and the
results of $\kappa$ as a function of $\rho$ for (a) $\Delta/t=1$,
(b) $\Delta/t=2$, (c) $\Delta/t=4$, and (d) $\Delta/t=6$ are plotted
in Fig. 4. In the case of hopping $t=0$, there are only three states
as shown in Fig.3: the Mott state at $\rho=1$, CDW state at
$\rho=1/2$, and the empty state at $\rho=0$, then zero
compressibility for all these three states is found. However, in the
present results shown in Fig. 4, we find that for the strong
alternating potential $\Delta$, zero compressibility is observed at
$\rho=1/2$, which show that CDW phase is an incompressible state.
Moreover, for all other densities except $\rho=0$ and $\rho=1$,
positive compressibility is observed, which indicates the absence of
the phase separation, and therefore confirms that the supersolid
state is the ground state of the system in this density range.

Now we give some physical interpretation to the above obtained
results. In analogy to the spinful Hubbard model \cite{varma}, the
hard-core boson Hubbard model (1) with an alternating potential
describes a competition between the kinetic energy ($\rho t$) and
alternating potential ($\Delta$). In the case of $\rho=0$, the
system is empty. With the increase of the number of hard-core bosons
($0<\rho<1$), there is a gain of kinetic energy per hard-core boson
proportional to $t$ due to the hopping. At the same time, CDW order
is reduced, costing an energy of approximately $\Delta$ per site.
The kinetic energy $\rho t$ favors the superfluid state and tends to
reduce CDW order, while the alternating potential $\Delta$ favors
CDW order and results in frustration of the kinetic energy, then
this competition results in the emergence of the supersolid phase.
For the weak alternating potential, the kinetic energy of the system
is larger than the potential energy, therefore CDW order is reduced
of several orders and the system is supersolid. With the increase of
potential energy, CDW order is enhanced, and in the region where the
potential energy of the system is larger than the kinetic energy,
CDW state appears at the hard-core boson density $\rho=1/2$. In the
case of $\rho=1$, both energy bands are completely occupied by
hard-core bosons, and therefore all hard-core bosons are localized.
In this case, there is no ``charge degree of freedom'' in the
system, which is a Mott insulator.

In conclusion, within the hard-core boson Hubbard model with an
alternating potential, we have employed QMC method to investigate
the ground-state physical properties of the cold-atom on a optical
lattice. Our results show explicitly that the supersolid phase
emerges due to the presence of the alternating potential. At
half-filling $\rho=1$, the system is a Mott insulator. For the weak
alternating potential, the supersolid state appears for the whole
range of hard-core boson densities $0<\rho <1$. However, for the
strong alternating potential, the supersolid state can be observed
for $0<\rho <1$ except $\rho =1/2$, where CDW state appears.

Experimentally, the 2D optical lattice has been realized by using
four beams of lasers \cite{blakie}. Moreover, the cold-atom on this
2D optical lattice has been studied experimentally. In particular, a
superlattice potential that is similar to the alternating potential
discussed in the present case has been realized experimentally by
using multiple wavelength laser beams \cite{speil,jsebby}.
Therefore, it is possible that our present results of the cold-atom
on a two-dimensional square optical lattice with an alternating
potential can be verified by further experiments.

\begin{acknowledgments}
This work was supported by the training program foundation for the
Talents and the foundation from Beijing Education Commission, the
National Natural Science Foundation of China under Grant Nos.
10847155 and 10774015, and the funds from the Ministry of Science
and Technology of China under Grant Nos. 2006CB601002 and
2006CB921300.
\end{acknowledgments}


\newpage

\begin{references}

\bibitem{banderson} B. Anderson and M. Kasevich, Science {\bf 282},
1686 (1998).

\bibitem{corzel} C. Orzel, A. K. Tuchman, M. Fenselau, M. Yasuda and
M. A. Kasevich, Science {\bf 291}, 2386 (2001).

\bibitem{omorsch} O. Morsch, J. H. Muller, M. Cristiani, D. Ciampini
and E. Arimondo, Phys. Rev. Lett. {\bf 87}, 140402 (2001).

\bibitem{djaksch} D. Jaksch, C. Bruder, J. I. Cirac, C. W. Gardiner and P.
Zoller, Phys. Rev. Lett. {\bf 81}, 3108 (1998).

\bibitem{lguidoni} L. Guidoni, C. Trich¡äe, P. Verkerk and G. Grynberg,
Phys. Rev. Lett. {\bf 79}, 3363 (1997).


\bibitem{ggbatouni} G. G. Batrouni and R. T. Scalettar, Phys. Rev. Lett.
{\bf 84}, 1599 (2000).

\bibitem{swessel} S. Wessel and M. Troyer, Phys. Rev. Lett. {\bf 95},
127205 (2005).

\bibitem{rgmelko} R. G. Melko, A. Paramekanti, A. A. Burkov, A.
Vishwanath, D. N. Sheng, and L. Balents, Phys. Rev. Lett. {\bf 95},
127207 (2005).

\bibitem{mboninsegni} M. Boninsegni and N. Prokofev, Phys. Rev. Lett.
{\bf 95}, 237204 (2005).

\bibitem{jygan} Jing-yu Gan, Yu-chuan Wen and Yue Yu, Phys. Rev. B
{\bf 75}, 094501 (2007).

\bibitem{dheidarian} D. Heidarian and K. Damle, Phys. Rev. Lett.
{\bf 95}, 127206 (2005).

\bibitem{psengupta} P. Sengupta, L. P. Pryadko, F. Alet, M. Troyer,
and G. Schmid, Phys. Rev. Lett. {\bf 94}, 207202 (2005).

\bibitem{npensrose} N. Penrose and L. Onsager, Phys. Rev. {\bf 104},
576 (1956).

\bibitem{ekim} E. Kim and M. H. W. Chan, Nature {\bf 427}, 225 (2004);
Science {\bf 305}, 1941 (2004).

\bibitem{tegami} T. Egami, S. Ishihara, and M. Tachiki, Science
{\bf 261}, 1307 (1993).

\bibitem{sandvik} A. W. Sandvik and J. Kurkijarvi, Phys. Rev. B
{\bf 43}, 5950 (1991); A. W. Sandvik, J. Phys. A {\bf 25}, 3667
(1992); A. W. Sandvik, R. R. P. Singh, and D. K. Campbell, Phys.
Rev. B {\bf 56}, 14510 (1997); A. W. Sandvik, Phys. Rev. B {\bf 59},
R14157 (1999); O. F. Syljuasen and A. W. Sandvik, Phys. Rev. E {\bf
66}, 046701 (2002).

\bibitem{elpollock}E. L. Pollock and D. M. Ceperley, Phys. Rev. B
{\bf 36}, 8343 (1987).

\bibitem{min} Min-Chul Cha and Ji-Woo Lee, Phys. Rev. Lett., {\bf
98}, 266406 (2007).

\bibitem{varma} P. W. Anderson, Science {\bf 235}, 1196 (1987); S.
Schmitt-Rink, C. M. Varma, and A. E. Ruckenstein, Phys. Rev. Lett.
{\bf 60}, 2793 (1988).

\bibitem{blakie}  P. B. Blakie, C. W. Clark, J. Phys. B {\bf 37}, 1391
(2004).

\bibitem{speil} S. Peil, J. V. Porto, B. Laburthe Tolra, J. M.
Obrecht, B. E. King, M. Subbotin, S. L. Rolston, and W. D. Phillips,
Phys. Rev. A {\bf 67}, 051603(R) (2003).

\bibitem{jsebby} J. Sebby-Strabley, M. Anderlini, P. S. Jessen, and
J. V. Porto, Phys. Rev. A {\bf 73}, 033605 (2006).

\end{references}
\end{document}